\documentclass[sigconf]{acmart}
\usepackage[linesnumbered,ruled,vlined]{algorithm2e}

\def\BibTeX{{\rm B\kern-.05em{\sc i\kern-.025em b}\kern-.08emT\kern-.1667em\lower.7ex\hbox{E}\kern-.125emX}}
    
\copyrightyear{2019}
\acmYear{2019}
\setcopyright{acmlicensed}
\acmConference[GoodTechs '19]{GoodTechs '19: 5th EAI International Conference on Smart Objects and Technologies for Social Good}{September 25-27, 2019}{Valencia, Spain}
\acmPrice{15.00}
\acmDOI{10.1145/1122445.1122456}
\acmISBN{978-1-4503-9999-9/18/06}

\begin{document}

%
% The "title" command has an optional parameter, allowing the author to define a "short title" to be used in page headers.
\title{On Extending the Wireless Communications Range of Weather Stations using LoRaWAN}

%
% The "author" command and its associated commands are used to define the authors and their affiliations.
% Of note is the shared affiliation of the first two authors, and the "authornote" and "authornotemark" commands
% used to denote shared contribution to the research.
\author{Ermanno Pietrosemoli}
\email{ermanno@ictp.it}
\orcid{0000-0001-7083-5319}
\affiliation{%
  \institution{The Abdus Salam International Centre for Theoretical Physics}
  \city{Trieste}
  \state{Italy}
}
\author{Marco Rainone}
\email{mrainone@ictp.it}
\orcid{0000-0001-5882-772}
\affiliation{%
  \institution{The Abdus Salam International Centre for Theoretical Physics}
  \city{Trieste}
  \state{Italy}
}
\author{Marco Zennaro}
\email{mzennaro@ictp.it}
\orcid{0000-0002-0578-0830}
\affiliation{%
  \institution{The Abdus Salam International Centre for Theoretical Physics}
  \city{Trieste}
  \state{Italy}
}

%
% By default, the full list of authors will be used in the page headers. Often, this list is too long, and will overlap
% other information printed in the page headers. This command allows the author to define a more concise list
% of authors' names for this purpose.
\renewcommand{\shortauthors}{Pietrosemoli, et al.}

%
% The abstract is a short summary of the work to be presented in the article.
\begin{abstract}
Consumer grade weather stations typically involve transmitting sensor information to a digital console that provides readouts of the data being collected. The wireless range of such system is confined to about 100 m, making their use limited to urban environments. We present a device that decodes the data being sent by the weather station and forwards them using the emerging LoRaWAN technology. We designed the device keeping in mind the peculiar conditions of Developing Countries, in particular the low cost and low power requirements. This allows interesting applications in the realm of disaster prevention and mitigation using a network of numerous weather stations.
\end{abstract}



%
% Keywords. The author(s) should pick words that accurately describe the work being
% presented. Separate the keywords with commas.

%
% A "teaser" image appears between the author and affiliation information and the body 
% of the document, and typically spans the page. 
%  \Description{Enjoying the baseball game from the third-base seats. Ichiro Suzuki preparing to bat.}

%
% This command processes the author and affiliation and title information and builds
% the first part of the formatted document.
\maketitle

\section{Introduction}
For disasters prevention and mitigation, it is very useful to install weather stations in remote areas, usually far from the site where the gathered data can be acted upon. For instance, for landslide mitigation, one would like to know the amount of precipitation in a mountain far away from where the catastrophic consequences might be felt. In this case, the cost of the remote units is quite relevant since they might need to be replaced often, and besides a cheaper unit is less attractive to thieves.
Although cellular service is widespread, coverage in sparsely populated areas is often lacking and the recurring subscription cost might prove too much of a burden in some cases.
According to Tessier\cite{Tess}, \textit{"Communications costs between RWISs (Road Weather Information Systems) and a central data repository can be a significant recurring cost for the systems"}.
Emerging LPWAN (Low Power Wide Area Networks) technologies like LoRa, on the other hand, are  specifically tailored to achieve greater range and lower  TCO (Total Cost of Ownership, the sum of the initial cost and the recurring operating expenses).

\begin{figure}[h!]
\centering
\includegraphics[width=8cm]{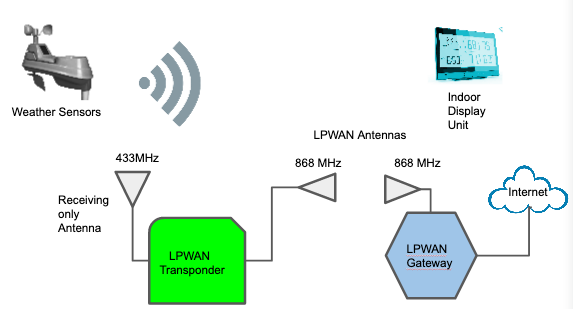}
\caption{The transponder captures the wireless signal sent by the weather sensors and forwards it to a LoRaWAN gateway.}
\label{fig1}
\end{figure}

LoRa has demonstrated the capability to span 300 km \cite{smart}, when there is an unobstructed path between transmitter and receiver. 
In this work, we present three versions of  a low cost "transponder",  shown in Figure~\ref{fig1}, that captures the radio signal used by many consumer grade weather stations (WSs) to convey the sensors' data to the indoor display unit. It then forwards them to the Network Server using the LoRaWAN architecture, via one or more Gateways. 

The design is open source and all the code we developed is freely available in Github \cite{git}  
By intercepting the wireless communication to the indoor device, there is no need to tamper with the Weather Station's sensors unit which could affect its calibration.

Weather data are used for a variety of purposes, according to the Weather Meteorological Organization (WMO) publication "ATLAS OF HEALTH AND CLIMATE" \cite{atlas} "...climate affects the geographical and temporal distribution of large burdens of disease and poses important threats to health security, .... climate information is now being used to protect health through risk reduction, preparedness and response over various spatial and temporal scales and in both affluent and developing countries"

Weather data from as many regions as possible are also required to build accurate weather prediction models and to test the validity of the ones currently in place. 
Professional weather stations are expensive, and although they have been widely deployed in affluent countries, few are operative in Africa and Latin America.
Mass produced consumer grade devices  cannot claim the same level of performance, but are  good enough for many applications. According to \cite{bell}, "...the high temporal and spatial resolution of freely available CWS (Citizens Weather Stations) data potentially lends itself to many applications".

The rest of the paper is organized as follows: Section 2  gives some background about weather stations. Section 3 provides a glimpse on LPWAN technologies. Section 4 describes the design of the transponder and in Section 5 we draw conclusions and hint at future work.

\section{Background}
\subsection{Automatic Weather Stations }

An Automatic Weather Station (AWS) comprises a number of meteorological sensors that automatically send their data to a remote server for storage, processing and visualization. The most common way of transmitting the data is by leveraging the cellular infrastructure that has been deployed all over the world. For developing countries, the high up front cost of professional Weather Stations and the recurring costs associated with cellular service have precluded their wide spread deployment, despite the recognition of the usefulness of the information provided. Furthermore, in sparsely populated areas the cellular phone coverage is lacking even in rich countries. 

The authors of \cite{bjorn} describe the building of an AWS in Uganda, but do not specifically address the issue of data transmission in absence of cellular infrastructure.
Although satellites can provide  global coverage, the recurring costs of the service make it currently unaffordable for AWS in developing countries.

\subsection{3D printed Automatic Weather Stations}

Recognizing the need for affordable weather monitoring in sparsely populated regions, the use of 3D printed plastic to build the mechanical parts coupled with low cost electronic platforms has been suggested. In particular, the University Corporation for Atmospheric Research (UCAR) and the US National Weather Service International Activities Office (NWS IAO) launched a program mired to develop observation networks and applications to reduce weather related risk. The units they developed offer the advantage that the locally printed parts can be easily replaced when needed. Their estimated cost of setting up a 3D AWS printing system is about 5000 USD and the cost of the materials for a printed unit are between 300 and 600 USD \cite{3dpaws}. Data is meant to be transmitted with a wireless adapter or a cell-modem setup, therefore they can also be transferred with a LoRa radio like the one we are proposing in order to achieve a wide coverage.

In our own experiments with 3D printed solutions, we observed significant differences in the readings of the outside temperature depending on the type of housing in which the sensor was inserted and even the color of the plastic used. This might lead to significant differences in the reported data. Similar conditions apply to the measurement of the wind speed, which are influenced by the type of bearing deployed. Although these issues can be solved by performing in situ calibration of the assembled unit, this is not easily doable in the field. On the other hand, the large scale manufacturing process of commercial weather stations implies replicability, so that units of the same type are identical and therefore can maintain a reasonably good calibration.

\subsection{Commercial LoRaWAN-based Weather Stations}

AWSs that use the LoRaWAN communication protocol are commercially available \cite{com} but their cost hinders their usage in developing countries. Furthermore, we could not find any information about calibration of such devices. The solution we propose has the advantage of decoupling the weather station and the long-distance transmission, allowing the use of pre-calibrated weather stations.

\subsection{Low Cost Personal Weather stations}

Numerous low cost personal weather stations can be found on the market \cite{Warne}. They are normally built with a split architecture, with the sensors unit placed outdoors to measure wind, rain, relative humidity and outside temperature, and the display unit placed indoor. As the atmospheric pressure is the same, often the barometer sensor is housed in the indoor unit. Some of the indoor units have an Ethernet connector so that data can also be transferred to a server on the cloud. Sensor's data are  transmitted to the indoor unit using unlicensed VHF frequencies reaching a maximum of 100 m. They range in cost from 100 to 800 USD and are readily available. 

We used stations made by La Crosse  and AcuRite, although the transponder developed could be adapted to many other brands.

\begin{figure}[h!]
\centering
\includegraphics[width=8cm]{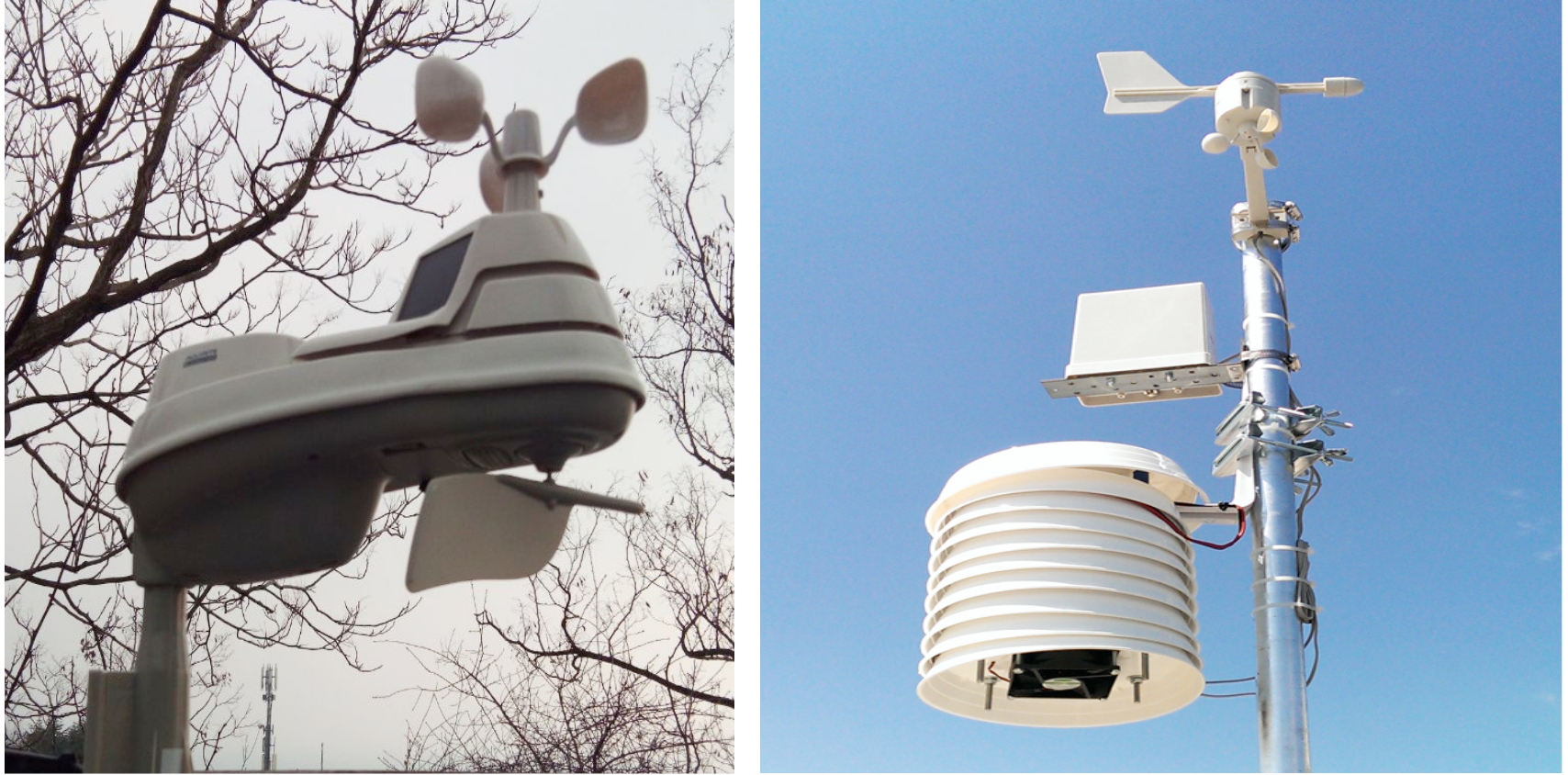}
\caption{Left: AcuRite outdoors integrated unit with anemometer, wind vane, rain gage, temperature and humidity sensors. Right: La Crosse  anemometer and wind vane at the top, rain gage in the middle and temperature and humidity sensors at the bottom.}
\label{fig2}
\end{figure}

\subsubsection{La Crosse WS-2300}

This WS \cite{LaCross} comprises an indoor unit, or "base station" with barometer, temperature and humidity sensors and a display that shows these data along with the one transmitted over the 434 MHz carrier by the outdoor sensors, with a maximum range in the open field of 25 m. The outdoor rain, temperature and humidity sensors are connected to the anemometer and wind vane by means of a 10 m long cable, allowing the installation of the wind sensor at 10 m above the ground, thus adhering to WMO \cite{WMOrec} recommendations for wind sensing.

This feature can be relevant for some applications and it is a differentiator from other WS in which all the sensors are included in a single housing. Furthermore, installing also our transponder at that height increases significantly the maximum range achievable by the LoRa transmission.

Since the transponder receives the data from the sensing unit by a wireless stream, it can be installed in any convenient location of its surroundings. The sensor unit uses 2 AA batteries which should last for approximately 12 months.

\subsubsection{AcuRite 5-in-1 Weather Station}

This \cite{acu} is a low cost WS that has been widely deployed and was used by the Massachusetts department of Transportation as a reference for the assessment of portable road weather information systems \cite{Tess}.  It comprises an outdoors sensors unit and an indoor display that can be up to 100 m away.
The integrated sensors unit has rain, wind speed, wind direction, humidity and temperature sensors. 
We added a barometric sensor to the trasponder unit as it is an important variable used in weather modeling, therefore we also send the barometric pressure over the LPWAN link. A barometer is included in the AcuRite indoor unit, but we will not use it in our application, since  the meteorological data will be  sent directly by the Transponder  to be displayed remotely.

The AcuRite has a built-in small photovoltaic panel used to power an internal fan that turns on when the temperature exceeds a certain threshold, its purpose is to improve he accuracy of the measurement by counteracting the effect of the heat accumulation  when the sun is stronger (thermal inertia).

The outdoor unit is powered by four AA batteries and transmits data over the unlicensed 433 MHz frequency. Figure~\ref{fig2} shows the  compact AcuRite outdoors unit on the right and the separate sensors of the La Crosse Weather Station on the left.

\begin{figure}[h!]
\centering
\includegraphics[width=8cm]{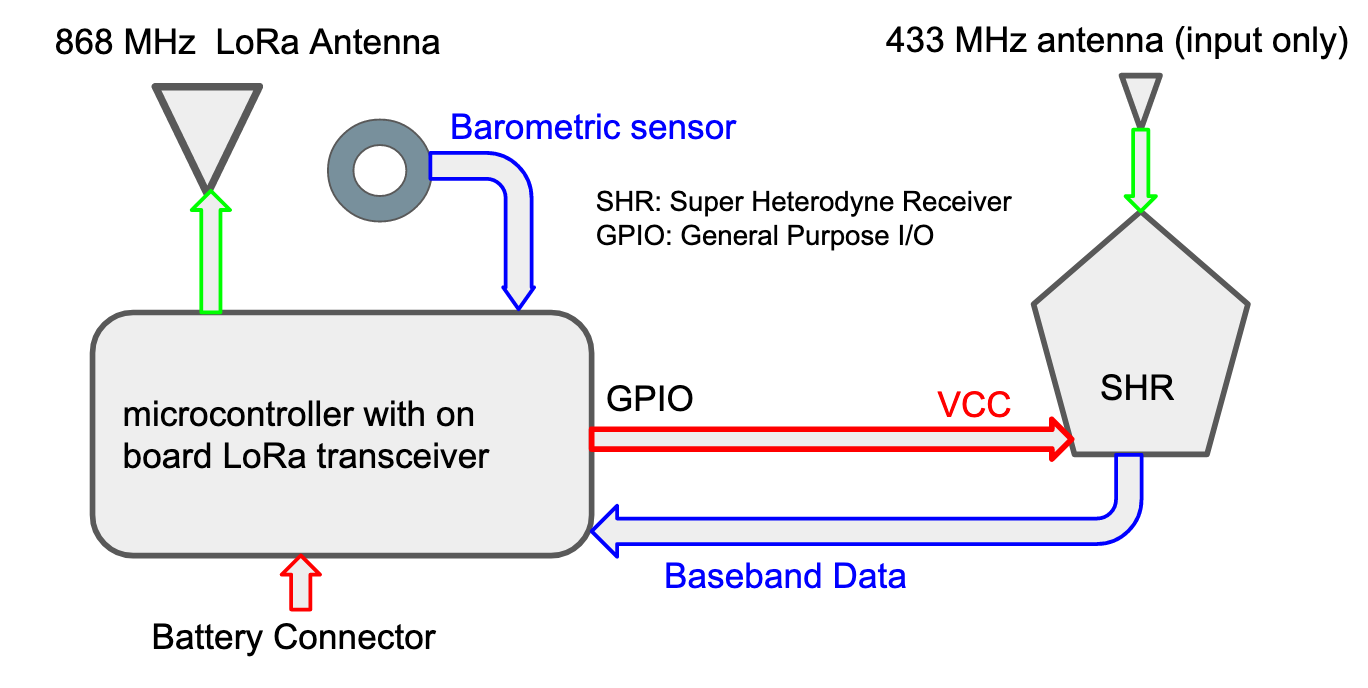}
\caption{Block diagram of the Transponder. The microcontroller with its transceiver and 868 MHz antenna, the barometer and the SHR with its 433 MHz wire antenna.}
\label{fig3}
\end{figure}

\section{Long Distance, Low Power Wide Area Networks}

\subsection{LoRaWAN}

Both SigFox or LoRaWAN can cover long distances by limiting the amount and rate of data transmitted. Weather data consist of a few bytes collected normally every 10 minutes, so they fit these requirements. Both solutions have been installed worldwide, but a significant difference is that SigFox service must be purchased from a commercial service provider whereas LoRaWAN offers also the possibility of autonomous installation by an interested party.

This opens up the possibility of deploying a LoRaWAN infrastructure even where no commercial service provider offers service, as was the case of Mozambique \cite{Salomao}, where a city wide LoRaWAN network comprising 10 Gateways has been installed in a very short time and at low cost.

A LoRaWAN infrastructure consists of a number of end nodes that connect to one or more Gateways, which in turn are normally connected to the Internet through which the required Network and Application servers are reached.  Several  commercial providers offer the service for a fee, but one can also install both the Network and Application server in a Unix box, even without internet access \cite{pape}.

The Things Network (TTN)\cite{TTN} is a community network that makes available its LoRaWAN infrastructure to anyone wishing to use it, at no cost. This means that one can install an end node, like the transponder that we designed, register it in the TTN web site to obtain the required credentials, and connect to an existing TTN Gateway and from there to the Internet.

The user traffic is encrypted all the way from the end node to the Application server in which the data can be processed according to the specific requirements.

In places where no TTN Gateway is available, one can still leverage the TTN's Network and Application servers by undertaking the installation of a GW. A number of LoRAWAN GWs are offered by many vendors, with prices that start at around 100 USD.

\section{Transponder Design and Implementation}

We designed the transponders following three design criteria:

\begin{itemize}
	\item \textbf{Low cost}: the trasponder should be affordable to allow for large deployments in Developing Countries. It should also be widely available on the market.
	\item \textbf{Low power}: as the battery life of the two WS we considered is about 2 years, the trasponder should have the same battery life expectancy. We considered sampling rates from 5 to 60 minutes to acommodate different scenarios.
	\item \textbf{Open source}: to facilitate tailoring the project to other applications, we used devices and software that adhere to the open source movement. 
\end{itemize}

\begin{figure}[h!]
\centering
\includegraphics[width=8cm]{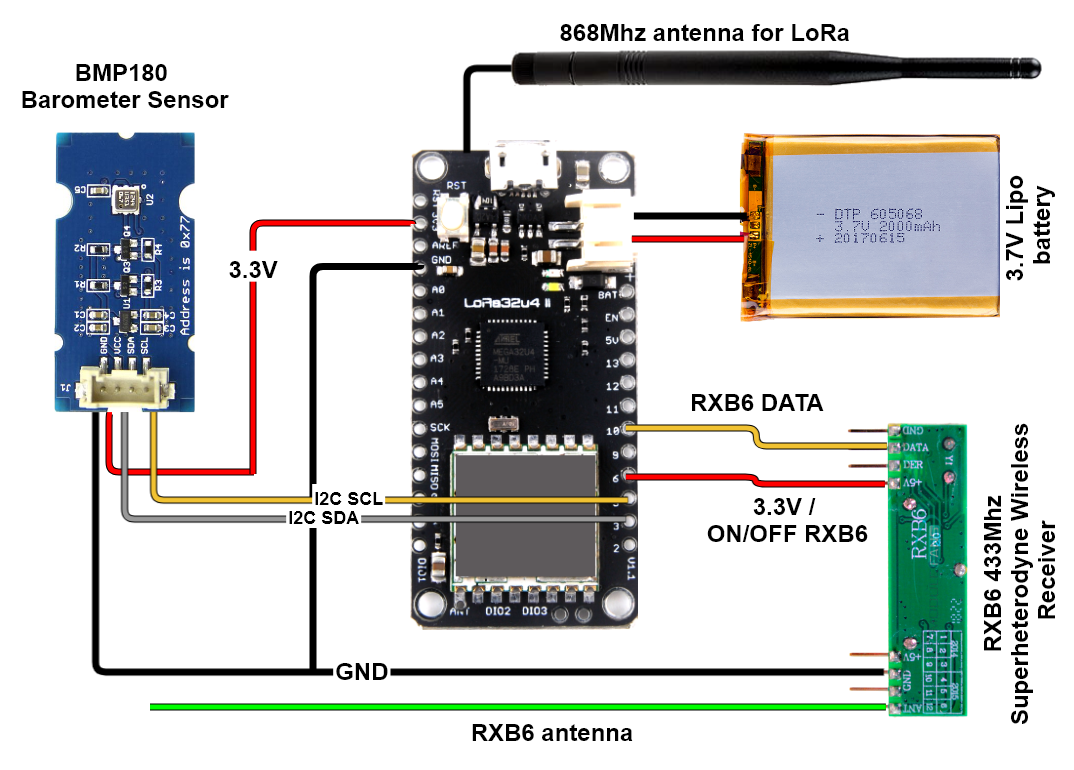}
\caption{BSF32 Transponder implementation.}
\label{fig4}
\end{figure}

\subsection{Decoding the transmitted signal}
We used on open source data receiver tool called rtl\_433 \cite{rtl} to analyze the data coming from the two different WSs. Once the data protocol was understood, we implemented the decoding in the microcontroller.

\subsection{LoRa Transponder}
The generic transponder, shown in Figure 3, intercepts the radio signals from the  sensing unit using the inexpensive RXB6 433 MHz super heterodyne receiver (SHR). A survey of the market revealed that the RXB6 433 was the cheapest receiver available in Europe that met our requirements.

We decoded the base band signal captured by the SRH with the microcontroller and sent it as the payload of a LoRaWAN frame to a TTN Gateway.  To save energy, the microcontroller turns off the SHR after finishing the reception of the data from the WS, and turns it back on when another frame is ready for processing. In Europe the frequency allocated to  LoRa transmissions lies in the  868 MHz band, for other countries the appropriate frequency should be configured.

The transponder uses two antennas, the one at 433 MHz is only for reception, so there is no need for precise matching and a simple piece of wire will do. The length of the wire is therefore not critical.

The 868 MHz LoRa antenna has an SMA connector, matched for optimum performance at 50 ohm, since it is used for transmission to the GW. The specific antenna deployed will depend on the topology. For a general purpose case an omnidirectional antenna with a small gain would be appropriate. For places in which there is a very long distance between the end node and the GW, a high gain directional antenna can be fitted to extend the coverage, thus meeting the requirements for remote weather stations monitoring in sparsely populated areas.

Figure 4 shows the  BSF32 Transponder implementation, on the left the barometer sensor, in the center the microcontroller equipped with the LoRa transceiver an its omnidirectional 868 MHz antenna, on the right the Li-ion battery above the super heterodyne receiver fitted with a simple wire as a 433 MHz receiving only antenna. 

For field deployment, the transponder is housed in an inexpensive waterproof enclosure made with RF transparent plastic that does not interfere with the antennas performance. The transponder can be installed at any convenient location, not too far from the WS (for instance, attached to the same pole), ideally with line of sight towards at least one of the LoRaWAN gateways. It will also work when the line of sight is blocked, but the range will be much shorter. 
 
As one of the key factors in designing the system was low power consumption, we measured the current absorption at different phases of the cycle using an OTII device \cite{otii}, specifically meant for  recording low current measurements with a very high time resolution. 

The microcontroller decodes the data received from the SHR and switches it off after reception, since the average consumption while using the SHR is 9.9 mA. 

In order to minimize the transmission air time in the ISM frequencies (which are normally constrained to a maximum duty cycle of 1\%), the string coming from the sensor's unit was  parsed to convey only the meteorological data, which are  encrypted according to the LoRaWAN protocol.

The decoded data are then passed to the SX1276 transceiver, which uses the LMIC-Arduino library to implement the LoRaWAN protocol required to connect to TTN. After completing transmission, the LoRa transceiver is switched off and the transponder enters the sleep phase during which the current is 144 uA. The time spent sleeping can be adapted to suit specific needs.  

\begin{figure}[h!]
\centering
\includegraphics[width=8cm]{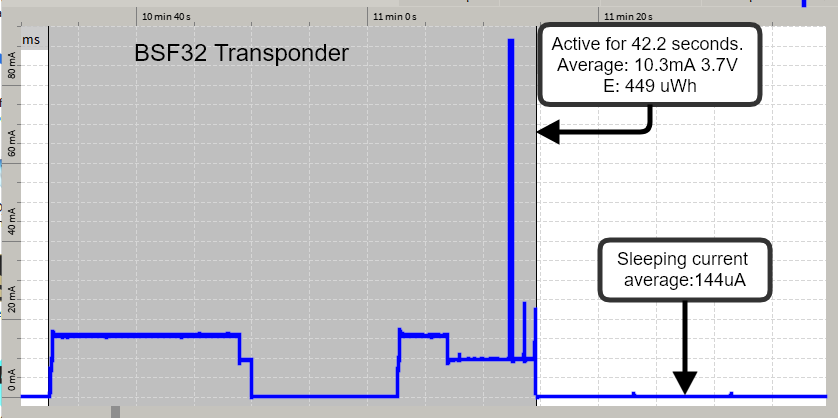}
\caption{Consumption versus time for the BSF32 transponder. The active phase, comprising receiving from the SHR, decoding, repacking the data and transmitting the LoRaWAN frame lasts 42.2 s. The duration of the sleeping phase, consuming 144 uA, will depend on the periodicity of transmission.}

\label{fig5}
\end{figure}

Figure~\ref{fig5} shows a graph of the measured current consumption at different stages for the whole transponder fed at 3.7 V. The active part lasts 42.2 s. The actual LoRa transmission consumes 102 mA but lasts only 289 ms when the 29 bytes payload is sent at spreading factor 9 and with an RF output power of 14 dBm. The total consumption will depend on the periodicity of transmission: for example, transmitting every 323 s (about 5 minutes), the consumption during the active phase is 449 uWh, while sleeping for 271 seconds consumes 40 uWh, for a total of 489 uWh per cycle.  The consumption per day is therefore 489X24X3600/323 = 130803 uWh, so a 3.7 V, 2000 mAh Li-ion battery will last for 56.7 days. 

In fact, we were able to run an experiment transmitting every 313 seconds and the 2000 mAh battery lasted 55 days. 

If the transmissions are done every 15 minutes, which is the recommended sampling rate for weather data, the consumption per cycle is 622 uWh, so per day it will be 622X24x4 = 59712 uWh, and the same battery will last 123 days. transmitting every 30 minutes will extend the battery duration to 204 days. 

This transponder was used with both the AcuRite and the  La Crosse WSs. Although their over the air protocol is different, the baseband was recovered by making appropriate changes in the coding.  The results were quite similar in terms of power consumption. One slight difference is that the La Crosse outputs 27 bytes per cycle instead of the 29 used by the AcuRite.

\begin{figure}[h!]
\centering
\includegraphics[width=8cm]{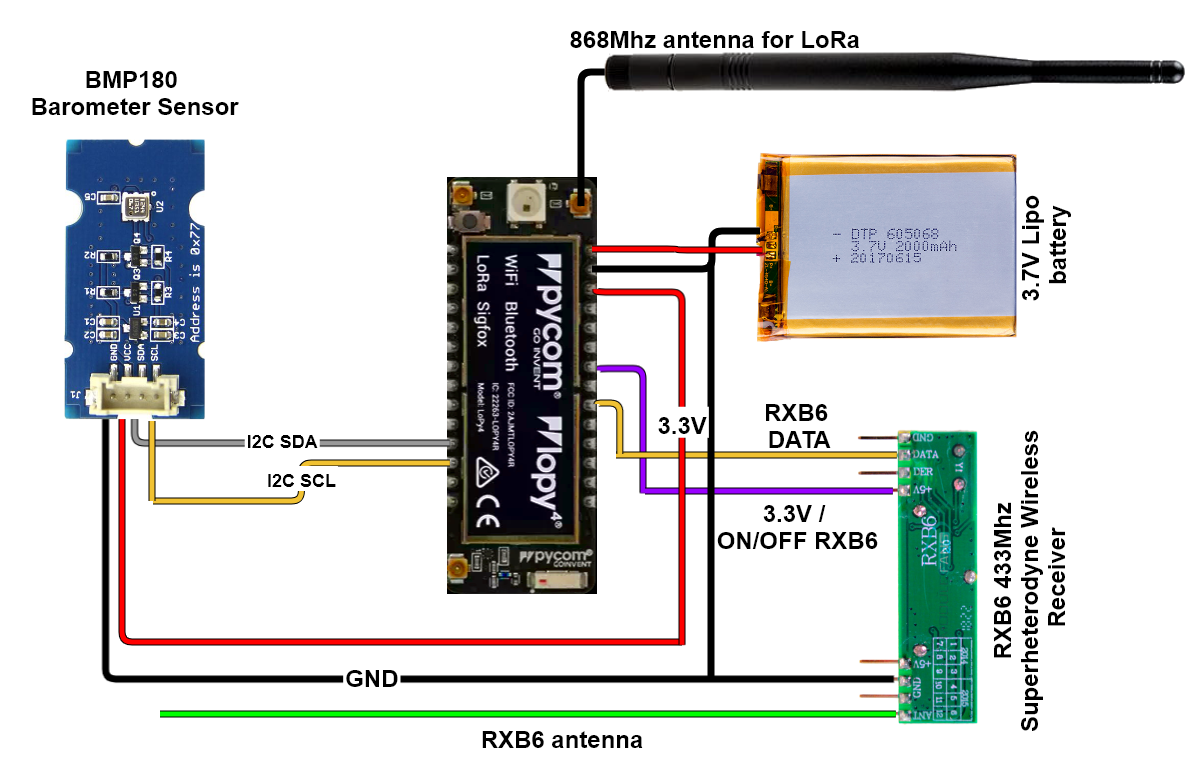}
\caption{Lopy4 Transponder. Left: Barometric sensor,
Right: RXB6 433 MHz receiver with wire antenna, Top: 868 MHz antenna.}
\label{fig6}
\end{figure}

One limitation of the BSF32 microcontroller is that to minimize the power consumption it must be fed through the JST (Japan Solderless Terminal) battery connector. If it is fed from the on-board micro USB connector, the power consumption is significantly higher since other circuit blocks are also activated.

Given that the project aim is to provide a solution with minimum end user involvement, we discarded the option to modify the feeding connector. This poses a problem since Li-ion batteries (that use the JST connector) are  not readily available in some markets, and they cannot be shipped due to current safety regulations. Therefore, after successfully testing this transponder with both of our weather stations, we decided to investigate another solution.

\subsection{Lopy4 Transponder}

\begin{figure}[h!]
\centering
\includegraphics[width=8cm]{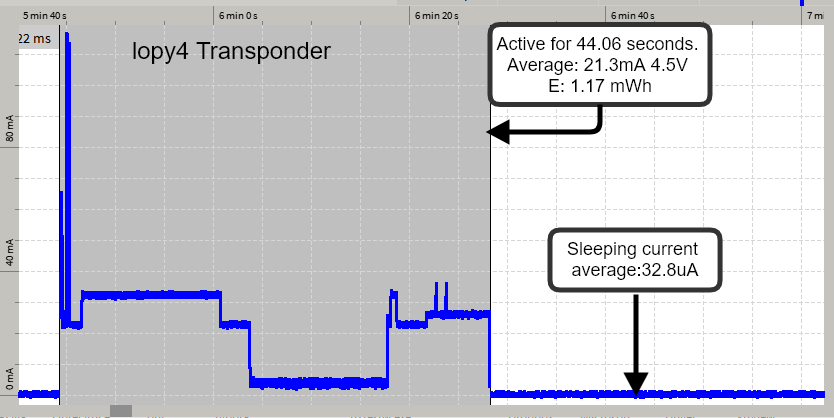}
\caption{Consumption versus time for the Lopy4 transponder. Active phase lasts 44.06 seconds, consuming 1.17 mWh. Sleeping current is 32.8 uA}
\label{fig7}
\end{figure}

A more versatile, albeit more expensive, version of the transponder was implemented using  the Lopy4 module, which is fitted with the powerful ESP32 processor and the same SX1276 LoRa transceiver. This microcontroller \cite{lopy} has more memory and a faster clock, which increases the consumption during the active phase. Nevertheless the sleeping current is only 32.8 uA and it can be powered with a voltage ranging from 3.5 to 5.5. This is a significant advantage as we can feed it with 3 type D alkaline batteries which are cheaper, widely available even in emerging economies, and have a total capacity of 48 Wh. Figure~\ref{fig7} shows the second transponder prototype. 

Even if  consumption is higher due to the powerful processor, the bigger battery pack allows for a longer overall duration. This is also a more versatile solution, given that besides the LoRa  it also has SigFox, WiFi and Bluetooth transceivers. 

Measurements with the OTII device showed that consumption in the active phase, which lasts 44.06 s, was 1.17 mWh, so when transmitting every 5 minutes  it spent 256 s in the sleep phase thus consuming 32.4uAX4.5VX256/3600h = 10.4 uWh, for a total of 1.18 mWh. In a day it will consume 1.18X12X24 = 33.98 mWh. A 48 Wh battery pack will therefore last 141 days.

If the transmission interval is extended to 15 minutes, the same battery pack will last 414 days. If we settle for transmission every half hour, the battery pack will last 739 days

Table 1 summarizes battery pack estimated duration for  the BSF32 fitted with a 7.4 Wh battery and  for the Lopy4 transponder fitted with a 48 Wh battery pack in four different scenarios of transmission frequency.

%Horizontal lines as row separators
%\begin{table}
%\begin{center}
%\small
%\caption{BSF32 Transponder Battery pack duration}
% \begin{tabular}{||c c||} 
% \hline
% BSF Transponder & 7.4 Wh battery \\
 % \hline\hline

% transmission freq., min & duration, days  \\ [0.5ex] 
% \hline\hline
% 5 & 56  \\ 
% \hline
% 15 & 123  \\
% \hline
% 30 & 204  \\
% \hline
%60 & 326  \\
% \hline\hline
 
% \end{tabular}
%\end{center}
%\end{table}

%Horizontal lines as row separators
%\begin{center}
%\small
%\caption{Lopy4 Transponder battery pack duration}
% \begin{tabular}{||c c||} 
% \hline
 %lopy4 Transponder, & 48 Wh battery \\
 % \hline\hline

% transmission freq., min & duration, days \\ [0.5ex] 
% \hline\hline
% 5 & 141 \\ 
% \hline
% 15 & 414 \\
% \hline
% 30 & 739 \\
% \hline
% 60 & 1478 \\
% \hline\hline
 
 %%\end{tabular}
%\end{center}
%\end{table}

\begin{table}
  \caption{Battery duration scenarios}
  \label{tab:freq}
  \begin{tabular}{ccc}
%    \toprule
     \hline
    transm. interval&BSF32 T. , 7.4 Wh batt.& Lopy4 T., 48 Wh batt.\\ 

 minutes&duration, days&duration, days\\    
 \hline

%    \midrule
    5&56&141\\
     \hline
  15&123&414\\
   \hline
   30&4204&739\\
    \hline
    60&326&1478\\
     \hline
 % \bottomrule
\end{tabular}
\end{table}%

%\subsection{Pseudo code of the Acurite 5n1 transponder management program}

% =========================================================================
\makeatletter
\newcommand{\nosemic}{\renewcommand{\@endalgocfline}{\relax}}% Drop semi-colon ;
\newcommand{\dosemic}{\renewcommand{\@endalgocfline}{\algocf@endline}}% Reinstate semi-colon ;
\newcommand{\pushline}{\Indp}% Indent
\newcommand{\popline}{\Indm\dosemic}% Undent
\makeatother

% =========================================================================
% pseudocodice algoritmo transponder

\begin{algorithm}
\DontPrintSemicolon % Some LaTeX compilers require you to use \dontprintsemicolon instead

\SetKwInput{KwReset}{RESET}
\SetKwInput{KwInit}{Initialize peripheral hardware}
\SetKwInput{KwConnect}{Connect to TTN}

\KwReset\;
\KwInit\;
\pushline
Bosch BMP180 sensor\;
RXB6 superheterodyne receiver\;
...
Initialize the LoRa interface\;
\popline
% \KwConnect\;
% \pushline
Connect to TTN. Setup LoraWan connection using ABP keys\;
% \popline
\BlankLine 
$T_{Cycle} \gets 15 minutes$\;
\While{$True$}{
	$T_{Start} \gets CurrentTime()$\;
    turn on the RXB6 receiver\;
    Enable the interrupt function to receive data from the RXB6 receiver\;
	\Repeat{$
		\textbf{the WS message is received completely}$}{
    	check status of the RXB6 receiver interrupt function\;
    }
    decode the first message\;
    update the data of the Weather Station currently in memory with new values read from the message\;
    \tcc{The next message will be received after at least ten seconds. 
    To save energy, go to sleep at least ten seconds.}
    Sleep(10 seconds)\;
	\Repeat{$
		\textbf{the WS message is received completely}$}{
    	check status of the RXB6 receiver interrupt function\;
    }
    decode the final message\;
    update WS data currently in memory with new values read from the message\;
    read the measurements from the BMP180 sensor: pressure and temperature\;
    turn off the RXB6 receiver\;
    form the message to be transmitted to TTN via LoraWan\;
	\Repeat{$
		\textbf{the LoraWan message has been completely transmitted}$}{
    	check lora status\;
    }
    \tcc{To save energy stay in deep sleep for the remaining time}
    turn off the RXB6 receiver and all unnecessary hardware\;
	$T_{End} \gets CurrentTime()$\;
	$T_{Active} \gets (T_{End} - T_{Start})$\;
	$TSleep \gets (T_{Cycle} - T_{Active})$\; 
    DeepSleep(TSleep)\;
}
% \Return{location}\;
\caption{{\sc Pseudo code of the Acurite 5n1 transponder management program}}
\label{algo:ws5n1Transponder}
\end{algorithm}

% end pseudocodice algoritmo transponder
% =========================================================================

\section{CONCLUSIONS}

Weather data are currently scarce in developing countries, despite their recognized relevance in many fields, not the least in disaster mitigation. To address this issue, we have designed and built inexpensive, low power and open source transponders that can capture the meteorological data from low cost commercial weather stations  and send them over long distances using the open LoRaWAN protocol. The code is freely downloadable from Github. 

Existing cloudsourced LoRaWAN TTN gateways can be leveraged to convey the gathered data to a target application server anywhere in the cloud at no cost.  Alternatively, inexpensive LoRaWAN Gateways can be installed by any interested party even in places where no Internet connectivity is available.  The LoRa reliance on unlicensed frequencies implies no recurring fee for data transmission. Our design does not interfere with the normal operation of the weather station, and the gathered data are stored in a  server accessible by any Internet enabled user device. 

We have tested two versions of the transponder based on the BSF32 microcontroller, to accommodate the needs of two types of commercial WSs, and a third version based on the more capable Lopy4. 

The latter  has additionally an on board SigFox transceiver, so it can also be used wherever SigFox networks are available by paying the corresponding subscription fee. Nevertheless the 12 bytes maximum payload of the Sigfox protocol will not accomodate the  size of the message transmitted by the weather stations, so some sort of data compression would need to be implemented. 

On the other hand, replacing the Lopy4 transponder with the Fipy model from the same manufacturer, which has also an NB-IoT transceiver, would allow  to leverage that infrastructure in places where NB-IoT service is available. This has the advantage of working in licensed bands, inherently protected from interference, and not subjected to the maximum air time limitations that affect both LoRa and Sigfox. 

Accurate consumption measurements reveal that in typical operation the batteries can last over a year, matching the replacement cycle of the weather sensor's batteries.  

Future work will address the installation of a number of Transponder's equipped weather stations to assess their performance in the field, leading to refinements of the design and its implementation in a single printed circuit that can be tailored to the output  protocols from different WSs by  a mere software configuration.

\end{document}